# Realistic large-scale modeling of Rashba and induced spin-orbit effects in graphene/high-Z-metal systems

*Elena Voloshina\* and Yuriy Dedkov\**

Prof. E. N. Voloshina,
Physics Department, Shanghai University, 99 Shangda Road, Shanghai 200444, P. R. China
Humboldt-Universität zu Berlin, Department of Chemistry, 10099 Berlin, Germany
E-mail: voloshina@shu.edu.cn

Prof. Y. S. Dedkov
Physics Department, Shanghai University, 99 Shangda Road, Shanghai 200444, P. R. China
E-mail: dedkov@shu.edu.cn

Graphene, as a material with a small intrinsic spin-orbit interaction of approximately 1 μeV, has a limited application in spintronics. Adsorption of graphene on the surfaces of heavy-metals was proposed to induce the strong spin-splitting of the graphene π bands either via Rashba effect or due to the induced spin-orbit effects via hybridization of the valence band states of graphene and metal. The spin-resolved photoelectron spectroscopy experiments performed on graphene adsorbed on the substrates containing heavy elements demonstrate the "giant" spin-splitting of the π states of the order of 100 meV in the vicinity of the Fermi level ($E_F$) and the K point. However, the recent scanning tunneling spectroscopy experiments did not confirm these findings, leaving the fact of the observation of the "giant" Rashba effect or induced spin-orbit interaction in graphene still open. Thus, a detailed understanding of the physics in such systems is indispensable. From a theory side this requires, first of all, correct modeling of the graphene/metal interfaces under study. Here we present realistic super-cell density-functional theory calculations, including dispersion interaction and spin-orbit interaction, for several graphene/high-Z-metal interfaces. While correctly reproduce the spin-splitting features of the metallic surfaces, their modifications under graphene adsorption and doping level of graphene, our studies reveal that neither "giant" Rashba- nor spin-orbit induced splitting of the graphene π states around $E_F$ take place.

## 1. Introduction

Spintronics is a field of science, the history of which can be traced back to the theoretical works of Rashba on the spin-orbit effects in non-centrosymmetric crystals[1] and to experimental works of Tedrow and Meservey on the tunneling in the ferromagnet/superconductor junctions,[2] dealing with a spin of electron and its influence on the transport properties of different materials and interfaces.[3-5] The progress in this field led to the discovery of the giant magnetoresistance effect in the metallic multilayered structures[6,7] followed by the spin-valve sensor allowing to increase the density of the data storage devices dramatically.[8,9] From the other side, the construction of field-effect transistor (FET), where channel conductivity is manipulated by the external electric field, and which is presently a main building block in the semiconductor technology, prompted the idea of the spin-FET device for the 2DEG (two-dimensional electron gas) in semiconductors, where spin-conductivity of the channel is also influenced by the electric field via the Rashba effect.[10,11]

Discovery of the unique properties of the truly 2D material graphene (gr) immediately places it in the forefront of spintronics research [12] with the idea to use it as a playground for different degrees of freedom of electron, like charge, (pseudo) spin, valley number, etc. Many spintronics applications on the basis of graphene were predicted and some of them were realized experimentally, like spin injection and transport on quite large distances of several μm.[13] However, the main dream on the spin manipulation in graphene via external electric field (Rashba effect) is not realized up to now due to the extremely small intrinsic spin-orbit interaction in carbon atoms (∼10 mK or ∼1 μeV[14]) and, respectively, very small spin-orbit induced gap of 24 μeV at the Fermi level ($E_F$).[15,16]

The first attempt for the spin-manipulation in epitaxial graphene was made in a decade-old angle-resolved photoelectron spectroscopy (ARPES) experiments on gr/Ni(111), where energy position of the graphene π states depends on the spin-direction of electrons in graphene and the energy shift for π states as large as 225 meV upon magnetization reversal of the system was detected.[17] This effect was explained by the induced spin-polarization of the graphene π states due to the underlying ferromagnetic Ni(111) substrate and by the asymmetry of the charge distribution at the graphene-metal interface as a result of the charge transfer from metal Ni 3d states to the graphene π states at the relatively short distance of ≈2.1 Å.[17] This initially observed large "Rashba-splitting" for the π states in gr/Ni(111) was under intensive discussion[18,19] and was finally confirmed by the observation of the induced spin-polarization of the graphene π states[20-23] as well as by the observation of the similar effect in gr/islanded-Co(0001) (see dispersion of the graphene-related π states for opposite magnetization directions in Ref. 19 with the maximal energy difference of ≈100 meV).

Furthermore, several methods were proposed for the increase of the spin-orbit interaction in graphene, like hydrogenation or fluorination.[24,25] Following Ref. 17, several works suggested adsorption of graphene on heavy-elements substrates with the idea of increasing the spin-orbit interaction in graphene via orbital mixing of the valence band states of graphene and the metallic substrate with high-$Z$.[26-31] These spin-resolved ARPES experiments demonstrate extraordinary "giant" energy splitting up to 100 meV of the graphene π states in the vicinity of the Fermi level and the K-point at the relatively large graphene-metal distances of 3.3-3.4 Å that generally should prevent the effective orbital hybridization between valence band states of graphene and metal. For example, the spin-splitting with the value of 13-100 meV for the graphene π states in the vicinity of $E_F$ and the K point was obtained for the same gr/Au/Ni(111) system in two different spin-resolved photoemission experiments by the same group.[26,27] These effects were ascribed to the induced strong spin-orbit interaction in graphene due to the hybridization of the graphene π states and Au 5d valence band states at large binding energies. (This effect of hybridization of the graphene and metal states at high binding energies was confirmed by spin-resolved and non-spin-resolved ARPES experiments).[26-28]   Similar effect of the π band spin-splitting of ≈40-60 meV was detected in spin-resolved ARPES experiments in the vicinity of $E_F$ and the K point for the gr/Ir(111) system.[29] Notable, high-resolution non-spin-resolved ARPES measurements performed on gr/Au and gr/Ir systems did not detect such splitting, although the measured graphene π bands are very "sharp" and clearly resolved.[32-37] Furthermore, the recent scanning tunneling spectroscopy (STS) measurements performed on graphene adsorbed on two metallic substrates demonstrating strong spin-orbit-induced splitting of the surface states, gr/Au(111) and gr/BiAg$_2$, rule out the observation of the "giant" spin-splitting of the graphene π states.[38,39] It is important to emphasize that the induced spin-splitting of the graphene π band, which originates from the hybridization of the graphene and metal valence band states, will strongly decrease when approaching $E_F$. At the same time the Rashba-splitting (if any) of these graphene π states will be constant as these states have the linear dispersion in the energy range of ±1 eV around $E_F$.

From the theoretical side it is important to note that experimental results on the observation of the "giant" spin-splitting of the graphene π states in the vicinity of $E_F$, and which are explained by the hybridization of the graphene π and Au 5d (or Ir 5d) states at high binding energies, are not supported by the density-functional theory (DFT) calculations, which for the realistic distance between graphene and Au(111) of 3.3 Å, yield a splitting value of only 9 meV.[27] Different computational tricks aiming to increase the spin-splitting of the graphene π-states lead either to the unrealistic distances between graphene and Au(111) of 2.3-2.5 Å and consequently to the incorrect doping level (+100 meV in experiment vs. -600 meV in DFT calculations) and binding energy of graphene on a metal substrate (which is smaller compared to the energy in the equilibrium geometry by 1eV), or to the crystallographic model, which is far away from the experimentally observed one.[27] The similar is valid for the recently published results for the other graphene/high-Z-metal interface, gr/Pb/Ir(111),[40] where noticeable effect on the graphene π states was obtained only for the artificially reduced distance between graphene and Pb layer of 2.7 Å instead of equilibrium distance of 3.33 Å obtained after system relaxation in DFT calculations (see further discussion in the text). As a consequence, the wrong value for the doping level of graphene was obtained in calculations (-750 meV in theory vs. -100 meV in experiment).[40]

Several of the above described works (e.g. Refs. 27, 40, 41) present the results of the DFT calculations used to support the experimental observations. However, these calculations are performed for the crystallographic models, which are restricted with respect to the real size of the real super-cells. As a consequence, the chosen lattice constants of graphene or metal surfaces are far from their equilibrium. Yet, graphene has very small density of states around $E_F$ in the zero-charge state and the small variations of the lattice constant of graphene or metal or distance between them leads to the dramatic modifications in the electronic structure of both parts. Therefore, such approaches cannot be considered as trustworthy in the studies of the graphene-based systems. Thus, the reliable non-compromising theoretical description of these systems, where all factors, like real system geometry, dispersive interaction and spin-orbit-interaction, are taken into account correctly, is required that can shed a light on the observed spin-related phenomena in graphene.

In this manuscript we present realistic super-cell DFT calculations of the spin-resolved electronic structure of graphene adsorbed on metallic substrates demonstrating the strong spin-orbit-induced splitting of the surface states, namely on Au(111), Ir(111), BiAg$_2$, Pb/Ir(111), Pb/Pt(111), and Au/Ni(111). All obtained results are compared with the available experimental data. Despite the correct description of the spin-resolved electronic structure of the metallic substrates before and after graphene adsorption (Rashba-splitting of the respective surface states) as well as the doping level of graphene, the "giant" spin-splitting of the graphene π states around $E_F$ and the K point of graphene is not reproduced.

## 2. Results and Discussion

Crystallographic structures of the systems under study are presented in **Table 1**: (a) gr/Ag(111), (b) gr/Au(111), (c) gr/Ir(111), (d) gr/BiAg$_2$, (e) gr/Au/Ni(111), (f) gr/Pb/Ir(111), and (g) gr/Pb/Pt(111). For all these graphene-based systems, due to the relatively large lattice mismatch between graphene and the underlying metal surfaces, the so-called surface moiré structures are formed as a result of different local atomic stackings.[42] In case of BiAg$_2$ and Pb/Ir(111) the $(\sqrt{3} \times \sqrt{3})R30°$ and $c(4 \times 2)$ commensurate with Ag(111) and Ir(111) structures are formed by the adsorbed heavy metals Bi and Pb at the graphene-metal interfaces, respectively.

The simplified band structure scheme of the studied systems is shown in **Figure 1** (example for the case of gr/Ag(111) or gr/Au(111)). Here two effects can be observed: (i) spin-splitting of the metallic substrate surface state around the Γ point due to the Rashba effect and (ii) possible spin-splitting of the graphene π states in the vicinity of the K point. The later effect can appear due to the possible hybridization with the valence band states of the underlying heavy metal or due to the Rashba effect. The respective values for the spin projections on the direction perpendicular to the wave vector direction k$_\parallel$ are color coded between red for +s$_y$ and blue for –s$_y$.

First, let us consider the modifications of the surface electronic structure of Ag(111), Au(111), and Ir(111) upon adsorption of graphene on these surfaces. For the clean metal surfaces listed above the well-known L-gap surface state around the Γ point, with the distinct Rashba splitting for Au(111) and Ag(111) (**Figure 2**, left column), is found in ARPES and STS experiments. For the clean metal surfaces the calculated binding energy of these states at the Γ point and spin-orbit energy splitting are: -0.132 eV and 0.015 eV at k$_\parallel$ = 0.12 Å$^{-1}$ for Ag(111), -0.57 eV and 0.165 eV at k$_\parallel$ = 0.18 Å$^{-1}$ for Au(111), and -0.305 eV and 0.245 eV at k$_\parallel$ = 0.06 Å$^{-1}$ for Ir(111), respectively. All these theoretical results are in very good agreement with data obtained in different spectroscopic experiments. [41,43-45]

Adsorption of graphene on close-packed surfaces of noble metals leads to the upward energy shift of the metal surface states with the conservation of the spin-topology and the energy splitting between spin-up and spin-down components for these states (Figure 2, middle column). The energy shifts for the surface states are: 0.28 eV for gr/Ag(111), 0.158 eV for gr/Au(111), and 0.163eV for gr/Ir(111). These shifts obtained theoretically are found to be in good agreement with the recent experimental results obtained by means of STS and ARPES [41,46] and can be explained by the stronger localization of the wave function of these states upon physisorption of graphene on metals compared to the clean metallic surfaces.

The presented supercell calculations also correctly reproduce the doping level of a graphene layer on different noble metal surfaces (Figure 2, right column). The calculated positions of the graphene Dirac point ($E_D$) in these systems are: $E_D - E_F$ = -0.65 eV for gr/Ag(111) and $E_D - E_F$ = 0.1 eV for gr/Au(111), which agree well with the experimental data (-0.56±0.08 eV and 0.10±0.02 eV for gr/Ag(111) and gr/Au(111), respectively).[46,47] In case of gr/Ir(111) the doping level of graphene as well as the opening of the energy gap at the K point for the graphene-derived π states is reproduced, which is a result of hybridization of these states and Ir surface state.[34,48]

We would like to emphasize that in the present work we do not restrict ourselves to smaller "cheaper" models for the considered graphene/metal interfaces in order to be absolutely sure that none of such restrictions might influence our results. For example the consideration of the $(2 \times 2)$-graphene lattice on $(\sqrt{3} \times \sqrt{3})R30°$-Au(111) in Ref. 46 leads to the strong variation of the doping level of graphene depending on the dispersion correction used. In case of a model used for the description of gr/Ir(111),[41] where graphene lattice was expanded to the lattice constant of Ir(111), the energy position as well as the energy shift of the Ir surface are not quantitatively reproduced, which casts doubt on the theoretically calculated doping level of graphene (although not presented in Ref. 41). *Therefore, the above presented above theoretical data for gr/Ag(111), gr/Au(111), and gr/Ir(111), demonstrating very good agreement between* *theory and experiment, confirm the correctness of the chosen approach used in the present work for the description of the graphene-metal interfaces.*

One of the intriguing findings for the graphene/heavy-metal interfaces, like gr/Au/Ni(111) or gr/Ir(111), is the experimental demonstration by means of spin-resolved ARPES of the extraordinary spin-splitting of the order of 100 meV for the graphene π states in the vicinity of $E_F$ and of the K point of the graphene-derived Brillouin zone and called as "giant" in Ref. 27. As was discussed in previous works, this effect was attributed to the hybridization of the heavy-metal d states and graphene π states. [26-28] Indeed, the effect of hybridization is observed for the energy ranges where d states are crossed by π states, that leads, e. g. in case of gr/Au(111), to the series of the avoid-crossed energy gaps in the band structure. The effect of hybridization at high binding energies is confirmed by our theoretical results as well (not shown). However, the calculated equilibrium distances for all considered graphene-metal interfaces are far too large leading to very weak hybridization between graphene π and heavy-metal d valence band states. As a result, a very small induced spin-orbit effect in graphene and consequently small spin-splittings of the graphene-derived states are obtained. The calculated mean distances between a graphene layer and the top metal layer are 3.085 Å for gr/Ag(111), 3.168 Å for gr/Au(111), and 3.392 Å for gr/Ir(111), respectively. Our results for distances between graphene and metal surfaces can be compared with the available experimental data for gr/Ir(111) which yield a value of 3.38 Å[49] giving very good agreement with our theoretical data and supporting our theoretical approach. These relatively large distances between graphene and high-Z metal support lead to the extremely small induced spin-orbit interaction for the graphene π electrons and, consequently, the small spin-splitting for the graphene π states is detected in the energy range of 1 eV below $E_F$ (Figure 2, right column). The resulting spin-splittings are summarized in **Table 2** and they clearly demonstrate the absence of the "giant" energy splitting between spin-up and spin-down components of the graphene π states in the vicinity of $E_F$ and the K point. For example it reaches sizable values of 15 meV at $E-E_F$ = -1.85 eV and k = 0.2 Å$^{-1}$ for gr/Ag(111), 50 meV at $E - E_F$

= -1.31 eV and k = 0.23 Å$^{-1}$ for gr/Au(111), 30 meV at E - E$_F$ = -0.73 eV and k = 0.14 Å$^{-1}$ for gr/Ir(111) (values of wave vector are given with respect to the K point). Moreover, as can be seen from presented results for gr/Ag(111), gr/Au(111), and gr/Ir(111) (Figure 2, right column and Table 2), the energy splitting between two spin components increases with the binding energy and reaches its maximum value where graphene π states intersect the metal d states. This result is clearly explained and confirmed by the effect of hybridization between graphene π and metal d states or/and by the relatively large graphene-metal distance of 3.2-3.4 Å that lead to the small induced spin-orbit interaction in graphene. This finding is opposite to the experimentally presented constant value of spin-splitting for π states shown in spin-resolved ARPES experiments.[26,27,29] The presented DFT calculations explain the absence of the observation of the spin-splitting for the graphene π states in the recent STS experiments. [38] *All our results presented here are self-consistent and give clear description of the considered graphene/high-Z metal interfaces.*

Similarly to the above presented cases we considered several more complex examples, such as gr/BiAg$_2$, and intercalation like systems gr/Au/Ni(111), gr/Pb/Ir(111), gr/Pb/Pt(111), which were intensively studied aiming to demonstrate the spin-orbit induced effects in graphene layers.[27,39,40,50] The crystallographic structures of these systems are shown in Table 1 (e-g).

The electronic structure of BiAg$_2$ surface alloy, which can be prepared on Ag(111), was intensively studied in the past decade.[51-53] In this surface alloy the Bi atoms form $(\sqrt{3} \times \sqrt{3})R30°$ structure with respect to Ag(111). This system demonstrates very big Rashba-splitting of the surface bands as observed in different spectroscopic experiments and confirmed theoretically (**Figure 3** (a)). Recently, the gr/BiAg$_2$ system was investigated with the aim at detection of a sizable spin-orbit induced phenomena in a graphene layer on the diluted layer of high-Z atoms.[39] It was shown that adsorption of a graphene layer on well-ordered BiAg$_2$ surface leads to the downward shift of the spin-resolved surface bands (these states are located around the Γ point of the Brillouin zone), i.e. the effect, which is opposite to the earlier considered case of the graphene adsorption on noble metal surfaces. This observation was confirmed by the respective DFT calculations and the effect was explained by the inward relaxation of Bi atoms in the Ag slab with the mean gr-Bi and gr-Ag distances of 3.277 Å and 3.90 Å, respectively, that leads to the delocalization of the wave function of the surface-related states [39] (Figure 3 (a,b)). At the same time, the spin topology of these bands almost does not change after graphene adsorption on the surface of BiAg$_2$.

The recent studies by means of STS of the gr/BiAg$_2$ system[39] show that no "giant" spin- splitting for the graphene π states at the K point was detected that confirmed by the present DFT calculations (energy of the spin-splitting is of the order of 10 meV, which is in the error bar of the present calculations; see Figure 3 (c)). The similar situation is also obtained for the gr/Au/Ni(111) system, where (9 × 9)gr/(8 × 8)Au(111) interface is formed (see Table 1 (e)). Our realistic DFT calculations for the fully relaxed system yield only 20 meV (at the energy which is by 160 meV below $E_F$) for the spin-splitting of the graphene π states in the vicinity of the K point (see Figure 3 (d)) and do not reproduce the experimentally obtained values of ≈100 meV.[27] Our present supercell DFT calculations for gr/BiAg$_2$ and gr/Au/Ni(111) correctly reproduce the doping level of graphene in these systems, i. e. *n* and *p*, respectively, with the correct position of $E_D$ for the π states (cf. Figure 3 (c,d) and data in Refs. 39, 47: gr/BiAg$_2$: -400 meV (experiment) vs. -590 meV (theory); gr/Au/Ni(111): +100 meV (experiment) vs. +50 meV (theory)) demonstrating again the self-consistency of the approach used in the present work. Here we would like to emphasize that the earlier presented models for the explanation of the experimentally observed "giant" spin-splitting of the graphene π states are either inconsistent (wrong crystallographic model, not correct description of the electronic structure, wrong doping level of graphene, etc.)[27] or not supported by the experimental data that leads to the unrealistic crystallographic structure and to the incorrect description of the electronic properties for these systems.[54]

Several recent experiments demonstrate the observation of phenomena, which were interpreted as the existence of the strong spin-orbit induced interaction in graphene after the interface was modified with heavy elements, namely with Pb.[40,50] In both works the observed energy level quantization (gr/Pb/Ir(111)[40]) as well as the gap opening at the Dirac point (gr/Pb/Pt(111)[50]) were interpreted as an indication of such an effect. However, the following works doubted the gap opening and the detected spin-polarization of the graphene π states as well as the interpretations. [55] Moreover, in order to reproduce the experimental data, the authors of Ref. 40 performed DFT calculations for ordered (2 × 2) arrays of Pb atoms adsorbed on isolated graphene with a distance between graphene and Pb of 2.7 Å. These model calculations with the artificial graphene-Pb distance yield the gap opening and also the spin-orbit-induced spin-splitting for the graphene π states. Note, however, the used distance between graphene and Pb and hence the obtained doping level of a graphene layer are far from reality.

We performed the realistic supercell modelling for both systems, gr/Pb/Ir(111) and gr/Pb/Pt(111), with and without inclusion of the spin-orbit coupling. The crystallographic model for these graphene intercalation-like systems is shown in Table 1 (f,g) and results are compiled in **Figure 4**. The mean equilibrium distance between a graphene layer and Pb atoms placed underneath is 3.332 Å and 3.306 Å for gr/Pb/Ir(111) and gr/Pb/Pt(111), respectively. Inclusion of the spin-orbit interaction in DFT calculations leads to the moderate decrease of the doping level of graphene compared to the case without spin-orbit coupling and the position of the Dirac points $E_D - E_F$ = -0.245 eV and $E_D - E_F$ = -0.185eV, respectively, and these values are in reasonable agreement with the experimental values.[40,50] However, the band structures for the both systems demonstrate the absence of the energy gap around the Dirac point as well as the absence of the large spin-splitting for the graphene π states (see Table 2). This effect is clearly attributed to the relatively large distance between graphene and Pb that prevents the appearance of the induced spin-orbit interaction in the graphene layer.

**3. Conclusions**

We performed the to date most realistic DFT studies of the different graphene-metal systems in the supercell geometry. The subject of our study is the claim of the experimental observation of the spin-orbit-induced "giant" spin-splitting (of the order of 100 meV) of the graphene π states in the vicinity of the Fermi level after its adsorption on the high-Z metals, like Au, Ir, or Pb. Several factors support the correctness of the employed computational approach in our work for the description of the graphene-metal interfaces: (i) all calculations are performed for the atomic geometries observed in the experiment thus considered structures model the real graphene-metal systems, (ii) our calculations correctly reproduce the experimentally measured graphene-metal distances (where experimental data are available), (iii) our calculations correctly describe, qualitatively and quantitatively, the electronic structure of metal surfaces before and after graphene adsorption (positions and shifts of surface states as well as values of the Rashba splitting), (iv) for all systems the doping level of graphene on metals agrees with available experimental data, (v) the electronic structure of graphene on metals is correctly described for a wide energy range and our theoretical data agree with experimental band structures. Our systematic studies demonstrate the absence of the "giant" spin-splitting of the order of 100 meV for the graphene π states in the vicinity of the Fermi level and around the Dirac point as stated in a series of the recent experimental works. The demonstrated theoretical value for such spitting does not exceed the value of 10-15 meV in the vicinity of $E_F$ and the K point of the graphene-derived Brillouin zone, that is in the error bar of the present large-scale calculations. The absence of the energy splitting between two spin components is argued by the relatively large distance of more than 3.2 Å between the metal support and the adsorbed graphene layer that leads to very weak hybridization between graphene π and heavy-metal valence band states. Therefore, we conclude that the present state-of-the-art one-electron calculations fully and correctly describe considered graphene-heavy-metal systems doubting the necessity of many-body relativistic considerations for these interfaces. Moreover, our findings keep open the validity of the recent experimental observations and limited theoretical approaches for the description of the graphene-based systems. The present results question the ability for graphene to be used for spin-manipulation in spintronics devices.

## 4. Experimental Section

*Computational details:* DFT calculations based on plane-wave basis sets of 500 eV cut-off energy were performed with the Vienna ab initio simulation package (VASP).[56,57] The Perdew-Burke-Ernzerhof (PBE) exchange-correlation functional was employed.[58] The electron-ion interaction was described within the projector augmented wave (PAW) method[59] with C (2s, 2p), Ni (3d, 4s), Ag (4d, 5s), Ir(5d, 6s), Pt(5d, 6s), Au (5d, 6s), Pb (6s, 6p), and Bi (6s, 6p) states treated as valence states. The Brillouin-zone integration was performed on Γ-centered symmetry reduced Monkhorst-Pack meshes using a Methfessel-Paxton smearing method of first order with $\sigma = 0.15$ eV, except for the calculations of total energies. For those calculations, the tetrahedron method with Blöchl corrections[60] was used. The k-mesh for sampling the supercell Brillouin zone are chosen to be as dense as at least $24 \times 24$, when folded up to the simple graphene unit cell. Dispersion interactions were considered by adding a $1/r^6$ atom-atom term as parameterized by Grimme ("D2" parameterization).[61] The spin-orbit interaction is taken into account.

The supercells used in this work are presented in Table 1. They are constructed from a slab of at least 5 layers of metal, a graphene layer adsorbed on one (top) side of a metal slab and a vacuum region of approximately 23 Å. The positions (x, y, z-coordinates) of C atoms, the ions of alloy (BiAg$_2$) or intercalant (Pb in gr/Pb/Ir(111) and gr/Pb/Pt(111) and Au in gr/Au/Ni(111)) as well as z-coordinates of the two topmost layers of a substrate were fully relaxed until forces became smaller than 0.02 eV Å$^{-1}$. The lattice constant in the lateral plane was set according to the optimized value of the respective bulk metal. When necessary, the bottom side of the slab was protected by a layer of Al atoms in order to break the inversion symmetry of the metal film.[62]

The band structures calculated for the studied systems were unfolded to the graphene (1 × 1) and the respective metallic substrate (1 × 1) primitive unit cells according to the procedure described in Refs. 63, 64 with the code BandUP.

**Acknowledgements**

The North-German Supercomputing Alliance (HLRN) is acknowledged for providing computer time.

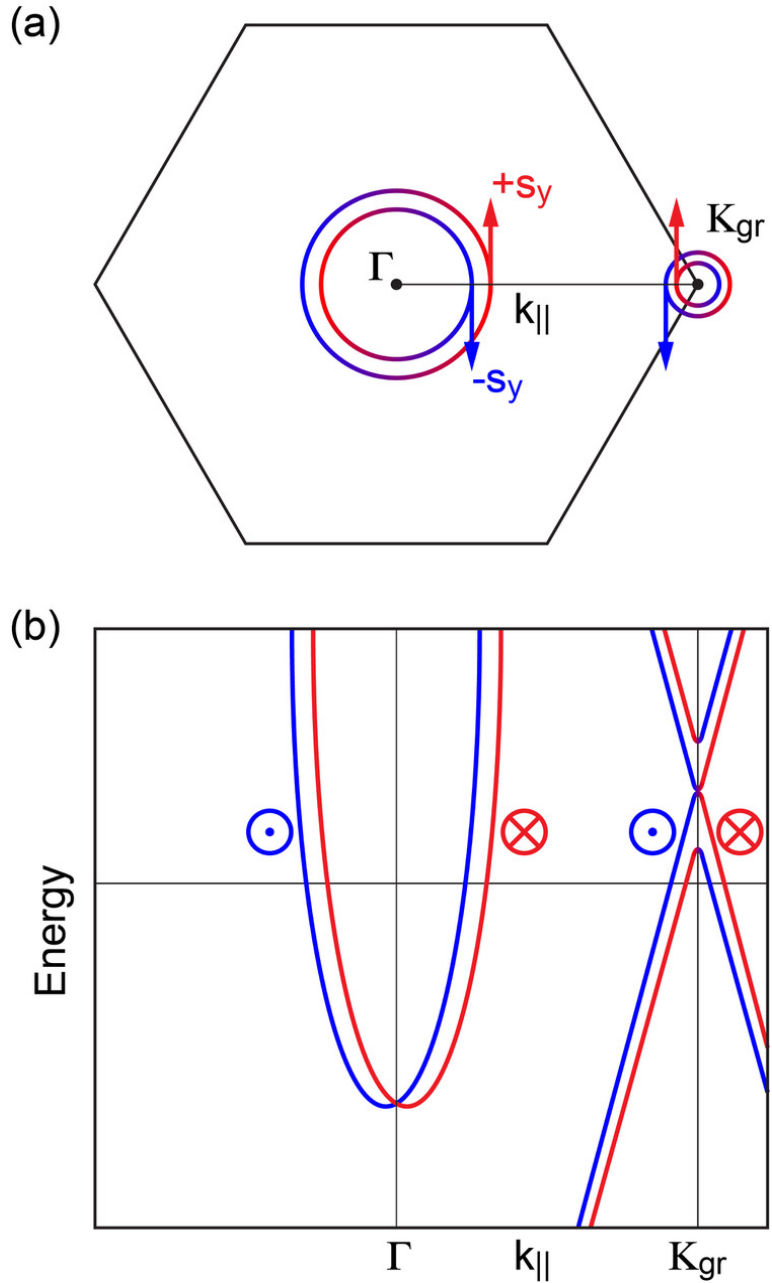

**Figure 1.** Example of the simplified band structure for the systems considered in the present manuscript (for gr/Ag(111) and gr/Au(111)) displaying the discussed features: (a) the constant energy cut and (b) the respective $E(k_{\parallel})$ plots for the Rashba-split surface state of metallic substrate (around the $\Gamma$ point) and graphene-derived $\pi$ states (around the $K_{gr}$ point). The value of the $\vec{s}$ projection on the direction perpendicular to $k_{\parallel}$ is color coded between red ($+s_y$) and blue ($-s_y$). The same color code is used in the following figures for the DFT calculated band structures.

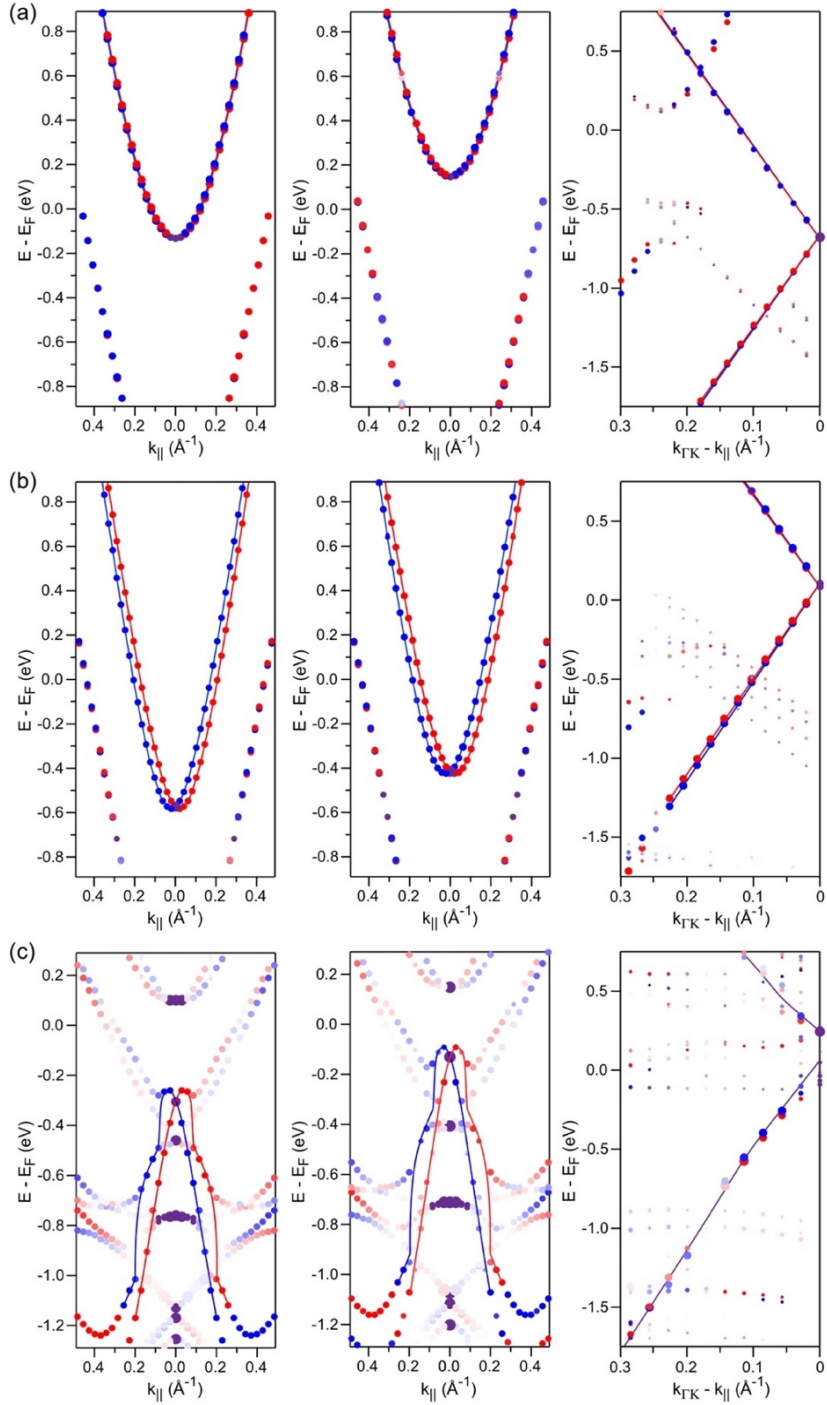

**Figure 2.** Spin-resolved band structures of different metal surfaces graphene-metal interfaces calculated around the Γ point and along the Γ − K direction of the graphene-derived Brillouin zone: (a) gr/Ag(111), (b) gr/Au(111), and (c) gr/Ir(111). Left and middle columns show data around the Γ point for the clean and graphene-covered metal surfaces. Right column show data along the Γ − K direction for the graphene-covered metal surfaces. Size of the filled circles gives the number of primitive cell bands crossing particular (k,E) in the unfolding procedure, i.e. the partial density of states at (k,E) for metal (left and middle columns) and graphene (right column), respectively.

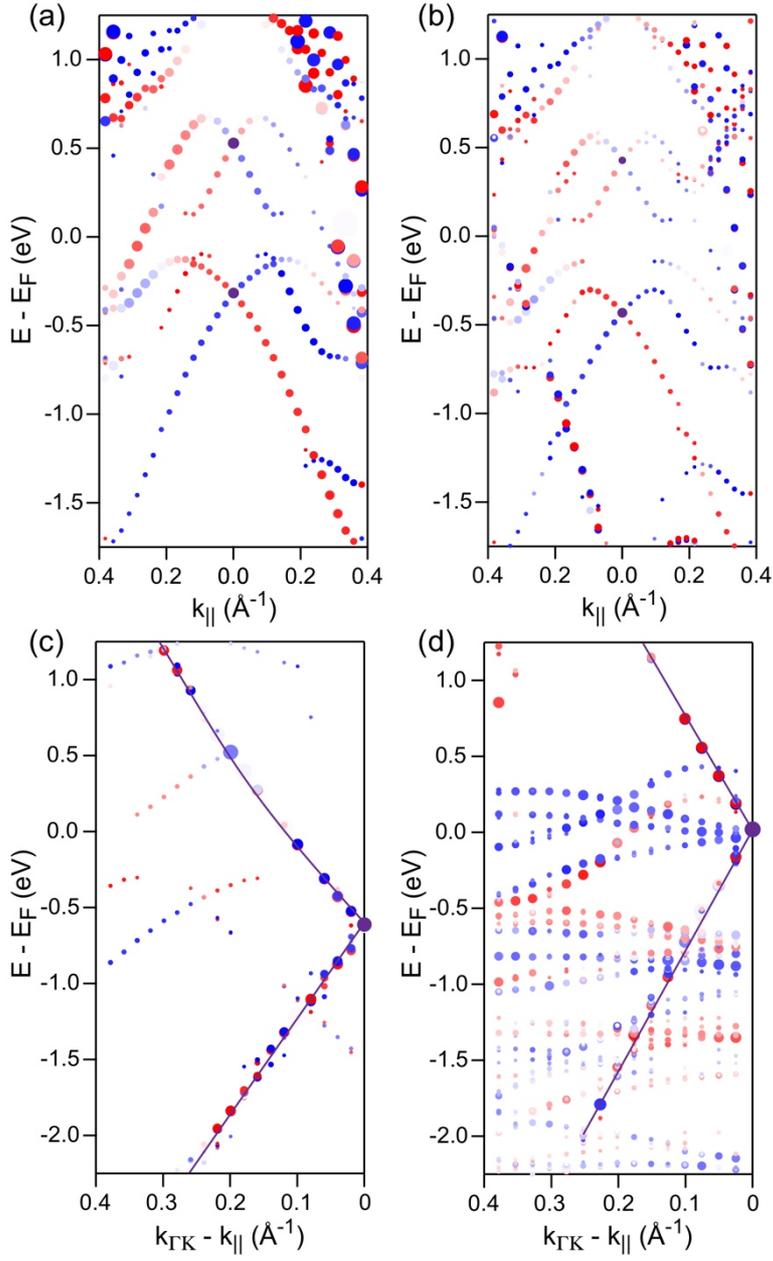

**Figure 3.** Spin-resolved band structures of different graphene-metal interfaces calculated around the Γ point: (a) and (b) gr/BiAg$_2$ before and after graphene adsorption and along the Γ−K direction of the graphene-derived Brillouin zone: (c) gr/BiAg$_2$ and (d) gr/Au/Ni(111). Size of the filled circles gives the number of primitive cell bands crossing particular (k,E) in the unfolding procedure, i.e. the partial density of states at (k,E) for metal (left and middle columns) and graphene (right column), respectively.

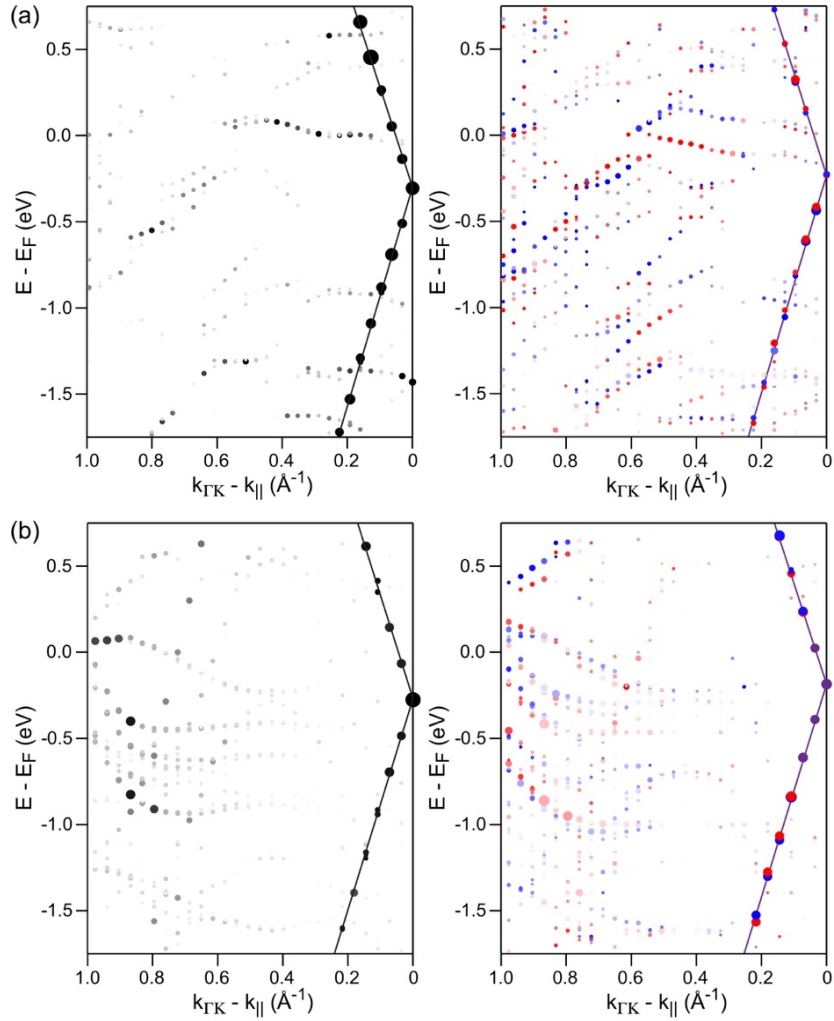

**Figure 4.** Band structures of different Pb-intercalation graphene-metal systems calculated along the Γ−K direction of the graphene-derived Brillouin zone: (a) gr/Pb/Ir(111) and (b) gr/Pb/Pt(111). Left- and right-hand columns present calculations without and with spin-orbit coupling, respectively. Size of the filled circles gives the number of primitive cell bands crossing particular (k,E) in the unfolding procedure, i.e. the partial density of states at (k,E) for metal (left and middle columns) and graphene (right column), respectively.

**Table 1.** Overview of the $(m \times m)$graphene/$(n \times n)$metal systems considered in the present work. Blue rhombi mark the studied supercells. The equilibrium distances obtained after geometry optimization are presented. Here $a_{gr}$ and $a_{M(111)}$ are the lattice constants of the (1×1) unit cells of graphene and the close packed metal surface M(111), respectively, d(C-M) is the mean distance between the graphene layer and the underlying metal layer (either surface or intercalated), d(Pb-Pb) is the shortest distance between the intercalated Pb atoms.

| System | Structure (top view) | Lattice parameters | Slab composition |
|---|---|---|---|
| (a) gr/Ag(111) | | $a_{gr}$ = 2.499 Å; $a_{Ag(111)}$ = 2.915 Å; d(C-Ag) = 3.085 Å; $m$ = 7; $n$ = 6 | C: 98 atoms; Ag: 216 atoms; Al: 36 atoms |
| (b) gr/Au(111) | | $a_{gr}$ = 2.422 Å; $a_{Au(111)}$ = 2.825 Å; d(C-Au) = 3.168 Å; $m$ = 7; $n$ = 6 | C: 98 atoms; Au: 216 atoms; Al: 36 atoms |
| (c) gr/Ir(111) | | $a_{gr}$ = 2.451 Å; $a_{Ir(111)}$ = 2.723 Å; d(C-Ir) = 3.392 Å; $m$ = 10; $n$ = 9 | C: 200 atoms; Au: 324 atoms; Al: 81 atoms |
| (d) gr/$(\sqrt{3} \times \sqrt{3})R30°$BiAg$_2$/Ag(111) | | $a_{gr}$ = 2.499 Å; $a_{Ag(111)}$ = 2.915 Å; $a_{BiAg2}$ = 5.049 Å; d(C-Bi) = 3.277 Å; $m$ = 7; $n$ = 6 | C: 98 atoms; Ag: 240 atoms; Bi: 12 atoms |
| (e) gr/Au/Ni(111) | | $a_{gr}$ = 2.458 Å; $a_{Ni(111)}$ = 2.458 Å; $a_{Au(111)}$ = 2.766 Å; d(C-Au) = 3.440 Å; $m$ = 9; $n$ = 9 | C: 162 atoms; Ni: 324 atoms; Au: 64 atoms; Al: 64 atoms |
| (f) gr/$c(4 \times 2)$Pb/Ir(111) | | $a_{gr}$ = 2.421 Å; $a_{Ir(111)}$ = 2.723 Å; d(Pb-Pb) = 4.717 Å; d(C-Pb) = 3.332 Å; $m$ = 9; $n$ = 8 | C: 162 atoms; Ir: 256 atoms; Pb: 16 atoms; Al: 64 atoms |
| (g) gr/$c(4 \times 2)$Pb/$(\sqrt{3} \times \sqrt{3})R30°$Pt(111) | | $a_{gr}$ = 2.413 Å; $a_{Pt(111)}$ = 2.786 Å; d(Pb-Pb) = 4.826 Å; d(C-Pb) = 3.306 Å; $m$ = 8; $n$ = 4 | C: 128 atoms; Pt: 192 atoms; Pb: 12 atoms |

**Table 2.** The resulting energy splitting between spin-up and spin-down components for the graphene-derived π states at different positions in the vicinity of the K point. Results are presented for all considered gr/heavy-metal systems.

| System | $k_\parallel - k_{\Gamma K}$ (Å$^{-1}$) | $E - E_F$ (eV) | Spin splitting (meV) |
|---|---|---|---|
| gr/Ag(111) | 0<br>0.10<br>0.20 | -0.65<br>-1.24<br>-1.85 | 0<br>10<br>150 |
| gr/Au(111) | 0<br>0.12<br>0.23 | 0.10<br>-0.65<br>-1.31 | 5<br>25<br>50 |
| gr/Ir(111) | 0.06<br>0.14 | -0.29<br>-0.73 | 20<br>30 |
| gr/BiAg$_2$/Ag(111) | 0.04<br>0.14 | -0.86<br>-1.43 | 15<br>10 |
| gr/Au/Ni(111) | 0.03<br>0.25 | -0.16<br>-1.55 | 20<br>60 |
| gr/Pb/Ir(111) | 0.03<br>0.26 | -0.42<br>-1.89 | 20<br>15 |
| gr/Pb/Pt(111) | 0.03<br>0.14 | -0.39<br>-1.06 | 0<br>5 |